\documentstyle[pra,aps]{revtex}
\begin{document}
\draft
\title{Disentanglement and Inseparability correlation : in two-qubit system}
\author{{\large Zheng-Wei Zhou\thanks{%
Email: lxli@mail.ustc.edu.cn} and Guang-Can Guo\thanks{%
E-mail: gcguo@ustc.edu.cn}}}
\address{Department of Physics and Lab. of\\
Quantum Communication and Quantum Calculation,\\
University of Science and Technology of China \\
Hefei 230026, P.R. China}
\maketitle

\begin{abstract}
Started from local universal isotropic disentanglement, a threshold
inequality on reduction factors is proposed, which is necessary and
sufficient for this type of disentanglement processes. Furthermore, we give
the conditions realizing ideal disentanglement processes provided that some
information on quantum states is known. In addition, based on fully
entangled fraction, a concept called inseparability correlation is
presented. Some properties on inseparability correlation coefficient are
studied.
\end{abstract}




\section{Introduction}

Entanglement describes a system composed of two or more particles, which
exhibits the astonishing property that the results of measurements on one
particle can not be specified independently of the parameters of the
measurements on the other particles. As one of the most striking features of
the quantum mechanics, entanglement is playing a more and more important
role in the young field of quantum information$\cite{Bennett1}$.

Recently, there has been growing interest in disentanglement of quantum
states. Usually an ideal disentanglement process can be described as:
transforming a state of two or more subsystems into an disentangled state (
generally a mixture of product states ) such that the reduced density
matrices of the subsystems are unaffected, i.e. 
\begin{equation}
\rho _{12}\longrightarrow \rho _{12}^{\prime }=\sum_i\omega _i\rho
_i^{\left( 1\right) }\otimes \rho _i^{\left( 2\right) }  \label{r1}
\end{equation}
with

\begin{eqnarray}
Tr_1\rho _{12} &=&Tr_1\rho _{12}^{\prime }\text{ ,}  \nonumber \\
Tr_2\rho _{12} &=&Tr_2\rho _{12}^{\prime }\text{ ,}  \label{r2}
\end{eqnarray}
where the positive weights $\omega _i$ satisfy $\sum_i\omega _i=1,$ and $%
\rho _i^{\left( 1\right) }$ and $\rho _i^{\left( 2\right) }$ refer to the
density matrices of the subsystem 1 and 2 respectively.

In an extreme case of disentanglement,both quantum entanglement and
classical correlations are eliminated, i.e., the final state becomes 
\begin{equation}
\rho _{12}\longrightarrow \rho _{12}^{\prime }=\rho ^{\left( 1\right)
}\otimes \rho ^{\left( 2\right) }.  \label{r3}
\end{equation}

It was recently shown by Terno $\cite{Terno}$ that an arbitrary state can
not be disentangled into a tensor product of its reduced density matrices (
i.e. to satisfy Eqs (\ref{r2}) and (\ref{r3})). Otherwise, in the process to
distinguish an quantum state secretly chosen from a certain set of two
entangled states, the probability that the observer makes a wrong guess will
be lower than Helstrom's minimal error probability, which can not be allowed
by quantum mechanics. Subsequently, Mor proved that an ideal universal
disentangling machine to satisfy Eqs (\ref{r1}) and (\ref{r2}) can not exist$%
\cite{Mor}.$ In other words, an arbitrary quantum entangled state can not be
disentangled. Similar to quantum no-cloning theorem$\cite{Wootters},$ this
result also stems from the linearity of quantum mechanics. Since ideal
universal disentanglement processes can not be realized, a natural question
is then what level of disentanglement can be reached by a universal
disentangling machine? Does there exist an optimal universal disentangling
machine? Started from Peres-Horodecki's separability criterion$\cite{Peres},$
S. Bandyopadhyay et. al have constructed a type of universal disentangling
machine by using local cloning operations$\cite{Band}.$ Afterwards, they
discussed optimal universal disentangling machines for two-qubit system and
obtained a result identical to the case of local cloning operations$\cite
{Ghosh}.$ In the present paper, we shall study the essential limits to the
universal processes of disentanglement under the condition of local
operations. In addition, arising from a class of special local operations,
the varies of correlation feature of the given composite system are also
considered.

In section 2, by analyzing the changes of Horodecki's T matrix induced by
local operations, a necessary condition for local universal disentangling
machines is derived. Furthermore, we prove that the necessary condition are
also sufficient for this type of disentanglement processes. We find that
Bandyopadhyay's results are only two special cases to satisfy this threshold
boundary. Moreover, for a certain set of two-qubit pure entangled states on
which one knows partial information, we put forward a necessary and
sufficient condition realizing ideal disentanglement by manipulating only
one part of states. In section 3, based on fully entangled fraction ( see
ref \cite{Bennett2}), a concept called inseparability correlation is
proposed and its properties under some local operations are studied. In the
last section, we give a brief argument on some aspects of disentanglement
and entanglement measure.

\section{Disentanglement by local operations}

\subsection{Local universal isotropic disentanglement processes}

Let $\rho ^{ent}$ be an entangled density matrix of two-qubit system; and $%
\rho _1$ and $\rho _2$ be the reduced density matrices of the subsystem 1
and 2, respectively. Then an universal isotropic disentanglement process is
defined as:

\begin{equation}
\rho ^{ent}\longrightarrow \rho ^{disent},  \label{r4}
\end{equation}
together with

\begin{eqnarray}
\rho _i^{\prime } &=&Tr_j\left( \rho ^{disent}\right) =\eta _i\rho _i+\frac{%
\left( 1-\eta _i\right) }2I,  \label{r5} \\
i &\neq &j\text{ ; }i\text{ }j=1\text{ }2\text{ ,}  \nonumber
\end{eqnarray}
where the parameter $\eta _i,$ called reduction factor, describes the
shrinking of the ith reduced density matrix. To meet the requirements of
universality in the disentanglement process, it is reasonable to require
that the reduced density matrices isotropically shrink. Due to the reduced
density matrices of the entangled system are mixed states, in general, the
fidelity $\cite{Uhlmann}$ between the reduced density matrices $\rho _i$ and 
$\rho _i^{\prime }$ : $F\left( \rho _i,\rho ^{\prime }\right) =\left( Tr%
\sqrt{\sqrt{\rho _i^{\prime }}\rho _i\sqrt{\rho _i^{\prime }}}\right) ^2$
can not remain a constant. Fidelity is no longer a suitable standard to
quantify of disentanglement. So, we use the reduction factors to define a
quality factor of disentanglement: 
\begin{equation}
Q=\frac{\eta _1+\eta _2}2\text{.}  \label{r6}
\end{equation}
when Q reaches its maximum, the corresponding universal disentanglement
process is optimal.

For a two-qubit system, it is convenient to use the Hilbert-Schmidt space
representation of the density matrices:

\begin{equation}
\rho _{12}=\frac 14\left( I\otimes I+\overrightarrow{r}\cdot \overrightarrow{%
\sigma }\otimes I+I\otimes \overrightarrow{s}\cdot \overrightarrow{\sigma }%
+\sum_{n,m=1}^3t_{mn}\sigma _m\otimes \sigma _n\right) ,  \label{r7}
\end{equation}
where $\rho _{12}$ acts on the Hilbert space $H=H_1\otimes H_2=C^2\otimes
C^2.$ $I$ stands for the identity operator, $\left\{ \sigma _n\right\}
_{n=1}^3$ are the standard Pauli matrices, and $\overrightarrow{r}$ and $%
\overrightarrow{s}$ are vectors in $R^3$ called Bloch Vectors with $%
\overrightarrow{r}\cdot \overrightarrow{\sigma }=\sum_{i=1}^3r_i\sigma _i.$
The coefficients $t_{mn}=Tr\left( \rho _{12}\sigma _m\otimes \sigma
_n\right) $ form a real matrix which we shall denote by $T\left( \rho
_{12}\right) $. Each part of reduced density matrices is the following:

\begin{equation}
\rho _1=Tr_2\rho _{12}=\frac 12\left( I+\overrightarrow{r}\cdot 
\overrightarrow{\sigma }\right) ,  \label{r8}
\end{equation}

\begin{equation}
\rho _2=Tr_1\rho _{12}=\frac 12\left( I+\overrightarrow{s}\cdot 
\overrightarrow{\sigma }\right) .  \label{r9}
\end{equation}

For any local general measurements, a particularly useful description is the
so-called operator-sum representation\cite{Kraus}:

\begin{equation}
\rho _1^{\prime }=\sum_iA_i\rho _1A_i^{+},  \label{r10}
\end{equation}

\begin{equation}
\rho _2^{\prime }=\sum_jB_j\rho _2B_j^{+}.  \label{r11}
\end{equation}
The two sets of operators satisfy the completeness relations $%
\sum_iA_i^{+}A_i=I$ and $\sum_jB_j^{+}B_j=I$. For simplicity of expression,
we use two completely positive and trace-preserving superoperators $%
\widetilde{V_i}$ ( i=1, 2) to characterize the above maps. In view of the
requirements of Eqs (\ref{r4}) and (\ref{r5}), the map relations can be
written as

\begin{equation}
\rho _1^{\prime }=\widetilde{V_1}\left( \rho _1\right) =\frac 12\left(
I+\eta _1\overrightarrow{r}\cdot \overrightarrow{\sigma }\right) ,
\label{r12}
\end{equation}

\begin{equation}
\rho _2^{\prime }=\widetilde{V_2}\left( \rho _2\right) =\frac 12\left(
I+\eta _2\overrightarrow{s}\cdot \overrightarrow{\sigma }\right) .
\label{r13}
\end{equation}
Due to the arbitrariness of the vectors $\overrightarrow{r}$ and $%
\overrightarrow{s}$ ( $\left| \overrightarrow{r}\right| ,\left| 
\overrightarrow{s}\right| <$ $1$ ), we can further deduce the map relations
satisfied by the identity operator and the Pauli operators: 
\begin{equation}
\left\{ 
\begin{array}{c}
\widetilde{V_i}\left( I\right) =I\text{ ,} \\ 
\widetilde{V_i}\left( \sigma _j\right) =\eta _i\sigma _j\text{ }\left( i=1,2;%
\text{ }j=1,2,3\right) .
\end{array}
\right.  \label{r14}
\end{equation}

Thus, after the joint action described by $\sum_{i,j}A_i\otimes B_j$, the
whole density matrix of the two-qubit system becomes

\begin{equation}
\rho _{12}^{\prime }=\frac 14\left( I\otimes I+\eta _1\overrightarrow{r}%
\cdot \overrightarrow{\sigma }\otimes I+\eta _2I\otimes \overrightarrow{s}%
\cdot \overrightarrow{\sigma }+\sum_{m,n=1}^3\eta _1\eta _2t_{mn}\sigma
_m\otimes \sigma _n\right) .  \label{r15}
\end{equation}

\subsection{Horodecki's tetrahedron}

For a long time, it is a puzzling question to check if a quantum system is
entangled or not. Peres and Horodecki et. al have studied this question$\cite
{Peres}.$ Horodecki et. al proved that Peres's criterion is necessary and
sufficient condition for separability of 2$\times $2 and 2$\times $3 systems.

For a two-qubit system, the correlations between subsystems are embodied in
the matrix $T\left( \rho \right) $\cite{Horodecki1}$:$

\begin{equation}
E\left( \overrightarrow{a},\overrightarrow{b}\right) =Tr\left( \rho 
\overrightarrow{a}\cdot \overrightarrow{\sigma }\otimes \overrightarrow{b}%
\cdot \overrightarrow{\sigma }\right) =\left( \overrightarrow{a},T%
\overrightarrow{b}\right) .  \label{r16}
\end{equation}

In accordance with Eq.(\ref{r7}), we can always select Pauli's operators in
a suitable representation to diagonalize the matrix $T$ :

\begin{equation}
T^{^{\prime }}=O_1TO_2^{+}=\left( 
\begin{array}{ccc}
t_1 & 0 & 0 \\ 
0 & t_2 & 0 \\ 
0 & 0 & t_3
\end{array}
\right) ,  \label{r17}
\end{equation}
where $O_i$ stands for a real orthonormal matrix. In this representation the
initial Bloch vectors $\overrightarrow{r}$ and $\overrightarrow{s}$are
changed into:

\begin{equation}
\overrightarrow{r^{^{\prime }}}=O_1\overrightarrow{r}\text{ , }%
\overrightarrow{s^{^{\prime }}}=O_2\overrightarrow{s}\text{.}  \label{r18}
\end{equation}
Then, the matrix $T$ is connected with the vector$\overrightarrow{t}=\left(
t_1,t_2,t_3\right) \left( \overrightarrow{t}\in R^3\right) .$ Horodecki et.
al have already systematically analyzed the space of the characteristic
vectors. For an arbitrary quantum state, all the characteristic vectors of
its matrix $T$ are contained in the tetrahedron $\widetilde{T}$ with
vertices $\overrightarrow{t_0}=\left( -1,-1,-1\right) ,$ $\overrightarrow{t_1%
}=\left( -1,1,1\right) ,$ $\overrightarrow{t_2}=\left( 1,-1,1\right) ,$ $%
\overrightarrow{t_3}=\left( 1,1,-1\right) .$ For all separable quantum
states, their characteristic vectors belong to the octahedron $\widetilde{L}$
with the vertices $o_1^{\pm }\left( \pm 1,0,0\right) ,$ $o_2^{\pm }=\left(
0,\pm 1,0\right) ,$ $o_3^{\pm }=\left( 0,0,\pm 1\right) $ (see Fig. 1). ( It
is worthy to stress that the vector $\overrightarrow{t}$ relevant to the
matrix $T$ is not unique because different orthonormal matrices can be
selected to diagonalize the matrix $T.$ It is easy to prove that in general
cases ($\left| t_1\right| \neq \left| t_2\right| \neq \left| t_3\right| $)
there are 24 characteristic vectors corresponding to its matrix $T$. )

Generally, not all the matrices $T$ are positive. For simplicity of
discussion, we introduce a positive matrix $M$

\begin{equation}
M=\sqrt{TT^{+}}.  \label{r19}
\end{equation}
Thus, the vector $\overrightarrow{t}=\left( t_1,t_2,t_3\right) $ is mapped
to the vector $\overrightarrow{m}=\left( \left| t_1\right| ,\left|
t_2\right| ,\left| t_3\right| \right) .$ At the same time, Horodecki's
tetrahedron $\widetilde{T}$ is mapped to the hexahedron $\widetilde{H}$ with
vertices $\overrightarrow{o}=\left( 0,0,0\right) ,$ $\overrightarrow{A}%
=\left( 1,0,0\right) ,$ $\overrightarrow{B}=\left( 0,1,0\right) ,$ $%
\overrightarrow{C}=\left( 0,0,1\right) ,$ $\overrightarrow{D}=\left(
1,1,1\right) $ ( see Fig. 2 ).

All the map relations are summarized in the following:

\begin{equation}
\left\{ 
\begin{tabular}{l}
$\overrightarrow{t}=\left( t_1,t_2,t_3\right) \longrightarrow 
\overrightarrow{m}=\left( \left| t_1\right| ,\left| t_2\right| ,\left|
t_3\right| \right) $ \\ 
$\widetilde{T}\longrightarrow \widetilde{H}$ \\ 
$\widetilde{L}\longrightarrow \widetilde{T}_{oABC}$%
\end{tabular}
\right. .  \label{r20}
\end{equation}

\subsection{The threshold condition for disentanglement}

For a 2$\times $2 composite system, people pay more attention to the
correlation between subsystems. C. H. Bennett et. al defined '' fully
entangled fraction'' of a density matrix $\rho $ \cite{Bennett2}, which can
act as an index to characterize nonlocal correlation: 
\begin{equation}
f\left( \rho \right) =\max \left\langle e\left| \rho \right| e\right\rangle ,
\label{a1}
\end{equation}
where the maximum is over all completely entangled states $\left|
e\right\rangle $. Fully entangled fraction can be indicated as another from 
\cite{Horodecki3}: 
\begin{equation}
f\left( \rho \right) =\frac 14\left( 1+N\left( \rho \right) \right) ,
\label{a2}
\end{equation}
where N($\rho $)=TrM($\rho $) ( see ref \cite{Horodecki1}\cite{Horodecki2}
). Although, relying on different representations, a density matrix $\rho $
normally can correspond to different matrix M($\rho $), the trace of matrix
M($\rho $) is unique.

In Fig 2, for all separable quantum states, the characteristic vectors of
their matrices M are contained in the tetrahedron $\widetilde{T}_{oABC}$,
while, all the dots in the tetrahedron $\widetilde{T}_{ABCD}$ correspond to
inseparable quantum states. We call $\widetilde{T}_{oABC}$ and $\widetilde{T}%
_{ABCD}$ separability and inseparability correlation region respectively.
Here, it is worthy to note that not all the quantum states with
characteristic vectors being located in $\widetilde{T}_{oABC}$ are
separable, but, as far as fully entangled fraction is concerned, these
states can not be different from separable quantum states.

$Q$ is an arbitrary dot in inseparability correlation region $\widetilde{T}%
_{ABCD}$ ( see Fig. 2 ). Line segment $oQ$ intersects plane $ABC$ at dot $%
Q^{\prime }.$ Let us define the inseparability correlation coefficient $I_c$
of the dot $Q$ :

\begin{equation}
I_c=\frac{oQ-oQ^{\prime }}{oQ^{\prime }}=N\left( \rho _Q\right) -1.
\label{r21}
\end{equation}
We stipulate that inseparability correlation coefficients of all the dots in
the region $\widetilde{T}_{oABC}$ are zero. Thereby, for any two-qubit
density matrix $\rho ,$ its inseparability correlation coefficient has the
following form:

\begin{equation}
I_c\left( \rho \right) =\left\{ 
\begin{array}{c}
0\text{ }\left( \text{ }N\left( \rho \right) \leq 1\text{ }\right) \\ 
N\left( \rho \right) -1\text{ }\left( \text{ }N\left( \rho \right) >1\text{ }%
\right)
\end{array}
\right. .  \label{r22}
\end{equation}
At a geometric intuitive angle, inseparability correlation coefficient
depicts the minimal distance from the characteristic vector of given two
qubits system to separability correlation region. In the hexahedron $%
\widetilde{H},$ all the dots in the identical plane perpendicular to the
orientation $\left( 1,1,1\right) $ have the same inseparability correlation
coefficient. The maximum of $I_c$ is in the dot $D,$ $I_{c(\max )}=2,$
corresponding to four Bell states.

By observing Eq(\ref{r15}), we find local universal isotropic disentangling
operations cause the matrix $T\left( \rho \right) $ shrink by the ratio $%
\eta _1\eta _2.$ In view of inseparability correlation coefficient $I_c,$ we
can achieve the necessary condition realizing local universal isotropic
disentangling operations:

\begin{equation}
\eta _1\eta _2\leq \frac 1{1+I_{c(\max )}}=\frac 13\text{.}  \label{r23}
\end{equation}
In the following part, we will prove that the necessary condition is also
sufficient for this type of disentangling operations.

For an arbitrary pure entangled state $\left| \Psi \right\rangle _i$,
without loss of generality, it can be indicated as $\left| \Psi
\right\rangle _i=\cos \theta \left| 00\right\rangle +\sin \theta \left|
11\right\rangle $ in a suitable representation. In this case we have: 
\begin{equation}
\rho ^i=\left( 
\begin{array}{cccc}
\cos ^2\theta & 0 & 0 & \sin \theta \cos \theta \\ 
0 & 0 & 0 & 0 \\ 
0 & 0 & 0 & 0 \\ 
\sin \theta \cos \theta & 0 & 0 & \sin ^2\theta
\end{array}
\right) .  \label{a3}
\end{equation}
We can use the coefficients of group operators in Hilbert Schmidt Space to
represent the above density matrix: 
\begin{equation}
\rho ^i=\frac 14\left( 
\begin{array}{cccc}
1+t_{33}+s_3+r_3 & 0 & 0 & t_{11}-t_{22} \\ 
0 & 1-t_{33}+s_3-r_3 & t_{11}+t_{22} & 0 \\ 
0 & t_{11}+t_{22} & 1-t_{33}-s_3+r_3 & 0 \\ 
t_{11}-t_{22} & 0 & 0 & 1+t_{33}-s_3-r_3
\end{array}
\right) .  \label{a4}
\end{equation}
( Here, we leave out the coefficients equal to zero.) These coefficients
satisfy the following relations: 
\begin{equation}
\left\{ 
\begin{array}{c}
s_3=r_3=s, \\ 
t_{11}=-t_{22}=t, \\ 
t_{33}=1, \\ 
\sum_{i=1}^3t_{ii}^2+s_3^2+r_3^2=3.
\end{array}
\right.  \label{a5}
\end{equation}
After a local universal isotropic disentangling operation to satisfy map
relation (\ref{r14}), the density matrix $\rho ^i$ is changed into: 
\begin{equation}
\rho ^o=\frac 14\left( 
\begin{array}{cccc}
1+\eta _1\eta _2+\left( \eta _1+\eta _2\right) s & 0 & 0 & 2\eta _1\eta _2t
\\ 
0 & 1-\eta _1\eta _2+\left( \eta _1-\eta _2\right) s & 0 & 0 \\ 
0 & 0 & 1-\eta _1\eta _2-\left( \eta _1-\eta _2\right) s & 0 \\ 
2\eta _1\eta _2t & 0 & 0 & 1+\eta _1\eta _2-\left( \eta _1+\eta _2\right) s
\end{array}
\right) .  \label{a6}
\end{equation}
We define matrix $\rho ^T$ as the partial transposition of $\rho ^o$ ( see
ref \cite{Peres}). Thus, $\rho ^T$ has the following form: 
\begin{equation}
\rho ^T=\frac 14\left( 
\begin{array}{cccc}
1+\eta _1\eta _2+\left( \eta _1+\eta _2\right) s & 0 & 0 & 0 \\ 
0 & 1-\eta _1\eta _2+\left( \eta _1-\eta _2\right) s & 2\eta _1\eta _2t & 0
\\ 
0 & 2\eta _1\eta _2t & 1-\eta _1\eta _2-\left( \eta _1-\eta _2\right) s & 0
\\ 
0 & 0 & 0 & 1+\eta _1\eta _2-\left( \eta _1+\eta _2\right) s
\end{array}
\right) .  \label{a7}
\end{equation}
Based on Peres-Horodecki criterion,the necessary and sufficient condition
satisfying $\rho ^o$ is separable follows that none of the eigenvalue of $%
\rho ^T$ is negative, i. e. : 
\begin{equation}
\lbrack 1-\eta _1\eta _2+\left( \eta _1-\eta _2\right) s][1-\eta _1\eta
_2-\left( \eta _1-\eta _2\right) s]-4\eta _1^2\eta _2^2t^2\geq 0.  \label{a8}
\end{equation}
Under the conditions (\ref{a5}), we can derive inequality: 
\begin{equation}
\left( 1-\eta _1^2\right) \left( 1-\eta _2^2\right) +\left[ \left( \eta
_1-\eta _2\right) ^2-4\eta _1^2\eta _2^2\right] t^2\geq 0.  \label{a9}
\end{equation}
when t reaches maximum ( $t_{\max }=1$), the threshold inequality for local
universal isotropic disentanglement processes is obtained: 
\begin{equation}
\eta _1\eta _2\leq \frac 13.  \label{a10}
\end{equation}
In addition, we know all the separable states still keep separable after
being performed local operations. Hence, we complete the proof.

S. Bandyopadhyay et. al have studied the disentanglement processes by using
local cloning operations \cite{Band}. Their main results can be summarized
as that it is possible to disentangle any two-qubit entangled state either
by applying local cloning on one of its qubits provided the reduction factor
of the isotropic cloner is less than or equal to $\frac 13,$ or by local
cloning of the individual subsystems provided the reduction factor of the
isotropic cloner used is less than or equal to $\frac 1{\sqrt{3}}.$
Obviously, the two kinds of cloning operations are only special cases to
satisfy the threshold condition (\ref{r23}) for local universal isotropic
disentangling operations.

At the first part of this section, we have defined the quality factor of
disentanglement $Q.$ Under local isotropic disentangling operations, based
on the threshold inequality (\ref{r23}), the maximal $Q$ can be obtained: $%
Q_{\max }=\frac 23,$ corresponding to the case $\eta _1=1,$ $\eta _2=\frac 13
$.

Finally, let us consider the following question: if the local universal
disentanglement processes satisfying the threshold limit (\ref{r23})
eliminate quantum entanglement of the given system, can they fully eliminate
classical correlations between subsystems and make the density matrix of the
whole system become the product of local density matrices? The answer is
negative unless $\eta _1\eta _2=0.$ It is shown in Fig 2 that for any pure
entangled state $\rho ,$ none of the eigenvalues of the matrix $M\left( \rho
\right) $ is zero (see ref \cite{account}). Hence, after the disentanglement
process each eigenvalue of the matrix $M\left( \rho \right) $ is only shrunk
by the ratio $\eta _1\eta _2$ ( $\eta _1\eta _2\neq 0$ ), while the
determinant of $M\left( \rho \right) $ is still nonzero. However, it is easy
to prove that a necessary condition for arbitrary quantum state with the
product form is that its matrix $M$ has zero determinant. As a result, when
the factor $\eta _1\eta _2$ is not equal to zero, local universal isotropic
disentangling operations can not completely cut off classical correlations
between subsystems.

\subsection{Ideal disentanglement in some special cases}

As we know, there are no universal ideal disentanglement processes. There
appears another question. Provided that some information about quantum
states is obtained, can we perform ideal disentangling operations for these
quantum states? Clearly, if we know all the information about an entangled
state, it can be disentangled completely. Suppose that Alice has a
biparticle system: $\left| \psi \right\rangle _{AB}=\cos \alpha \left|
0\right\rangle _A\left| 1\right\rangle _B+\exp \left( i\vartheta \right)
\sin \alpha \left| 1\right\rangle _A\left| o\right\rangle _B.$ She sends
particle B to Bob. If Bob has known the wavefunction of the biparticle state
and which particle is sent, he can prepare a pair of entangled particles
identical to Alice's: $\left| \Psi \right\rangle _{A^{\prime }B^{\prime
}}=\cos \alpha \left| 0\right\rangle _{A^{\prime }}\left| 1\right\rangle
_{B^{\prime }}+\exp \left( i\vartheta \right) \sin \alpha \left|
1\right\rangle _{A^{^{\prime }}}\left| 0\right\rangle _{B^{^{\prime }}}.$
Bob swaps particle B with particle B', and traces out the particles B and
A'. As a result, the state of the remaining particles is a product of mixed
states. This is a trivial result.

In general, we do not know all the information on quantum states. Assume
that Alice has a set of known two-qubit pure entangled states $S=\left\{
\left| \Psi \right\rangle _{AB}^{\left( i\right) }\right\} _{i=1}^n.$ She
randomly singles out quantum states from this set and transmits part B of
these states. Here, the reduced density matrices of different states satisfy
the relation $\rho _B^i\neq \rho _B^j$ ( i $\neq $ j ). Provided that a
particle B is captured by eavesdropper Eve in its transmitting process, in
what case Eve can perform an ideal disentangling operation for this
entangled state by only manipulating the particle B? To solve this question,
here, we introduce the following theorem:

Theorem 1 : The state secretly chosen from the set of two-qubit pure
entangled states $S=\left\{ \left| \Psi \right\rangle _{AB}^{\left( i\right)
}\right\} _{i=1}^n$ can be perfectly disentangled by only manipulating the
particle B if the reduced density matrices of the part B of all the states
in this set commute with each other, i.e. $\left[ \rho _B^i,\text{ }\rho
_B^j\right] =0$ ( i $\neq $ j ; i, j = 1.....n ).

Proof : If all the reduced density matrices $\rho _B^i$ commute with each
other, their spectral decompositions have the form:

\begin{equation}
\rho _B^i=\lambda ^i\left| s\right\rangle \left\langle s\right| +\left(
1-\lambda ^i\right) \left| s_{\bot }\right\rangle \left\langle s_{\bot
}\right| ,  \label{r24}
\end{equation}
where $\left\{ \left| s\right\rangle ,\left| s_{\bot }\right\rangle \right\} 
$ is a set of orthonormal basis for the qubit system. One can use the
particle B as the control bit and an additional qubit as the target to
perform a CNOT ( Controlled NOT ) operation:

\begin{equation}
\left\{ 
\begin{array}{c}
\left| s\right\rangle \left| 0\right\rangle _c\longrightarrow \left|
s\right\rangle \left| 0\right\rangle _c \\ 
\left| s_{\bot }\right\rangle \left| 0\right\rangle _c\longrightarrow \left|
s_{\bot }\right\rangle \left| 1\right\rangle _c
\end{array}
\right. ,  \label{r25}
\end{equation}
where $\left| 0\right\rangle _c$ and $\left| 1\right\rangle _c$ are
orthonormal states. After taking the trace over the additional qubit, $\rho
_{AB}^{\left( i\right) }$ becomes a mixture of product states, while its
reduced density matrices keep invariant, i. e. a perfect disentanglement
process is fulfilled.

In the special cases that the eavesdropper only knows some information on
the state set, we get another theorem:

Theorem 2 : For a set of two-qubit pure entangled states $S=\left\{ \left|
\Psi \right\rangle _{AB}^{\left( i\right) }\right\} _{i=1}^n,$ the reduced
density matrices of the part B are assumed to have the form $\rho
_B^i=\lambda ^i\left| s^i\right\rangle \left\langle s^i\right| +\left(
1-\lambda ^i\right) \left| s_{\bot }^i\right\rangle \left\langle s_{\bot
}^i\right| .$ If one only knows the orthonormal basis of $\rho _B^i$'s
spectral decompositions $\left\{ \left| s^i\right\rangle ,\left| s_{\bot
}^i\right\rangle \right\} _{i=1}^n$ but not spectral coefficients $\left\{
\lambda ^i\right\} _{i=1}^n,$ he can perform an ideal deterministic
disentangling operation for this set by only manipulating the particle $B_i$
if and only if $\left[ \rho _B^i,\text{ }\rho _B^j\right] =0$ ( i $\neq $ j
; i, j = 1.....n ).

Proof : The sufficient condition can be easily obtained from the theorem 1.
Our task remains to prove the converse.

Suppose that there are two density matrices $\rho _B^i$ and $\rho _B^j$ with 
$\left[ \rho _B^i,\text{ }\rho _B^j\right] \neq 0.$ We introduce an
additional system C in the state $\left| e_0\right\rangle .$ (if the state
of the system is mixed, without loss of generality, we can view it as a pure
state in a larger space). Then, a joint unitary evolution between the
particle $B_i$ and the additional system is performed:

\[
\left| \Psi \right\rangle _{AB}^{\left( i\right) }\left| e_0\right\rangle
\longrightarrow 
\]

\begin{equation}
e^{i\Delta _1^{\left( i\right) }}\sqrt{\lambda ^i}\left| s^{iA}\right\rangle
U_{BC}\left( \left| s^{iB}\right\rangle \left| e_0\right\rangle \right)
+e^{i\Delta _2^{\left( i\right) }}\sqrt{1-\lambda ^i}\left| s_{\bot
}^{iA}\right\rangle U_{BC}\left( \left| s_{\bot }^{iB}\right\rangle \left|
e_0\right\rangle \right) .  \label{r26}
\end{equation}

An ideal disentanglement process requires that the reduced density matrices
of the original state remain unchanged, i.e.,

\begin{equation}
\rho _A^i=Tr_{BC}(\rho _{ABC}^i)=\lambda ^i\left| s^{iA}\right\rangle
\left\langle s^{iA}\right| +\left( 1-\lambda ^i\right) \left| s_{\bot
}^{iA}\right\rangle \left\langle s_{\bot }^{iA}\right| ,  \label{r27}
\end{equation}

\[
\rho _B^i=Tr_C\left( Tr_A\left( \rho _{ABC}^i\right) \right) 
\]

\[
=Tr_C\left\{ \lambda ^iU_{BC}\left( \left| s^{iB}\right\rangle \left|
e_0\right\rangle \left\langle e_0\right| \left\langle s^{iB}\right| \right)
U_{BC}^{+}+\left( 1-\lambda ^i\right) U_{BC}\left( \left| s_{\bot
}^{iB}\right\rangle \left| e_0\right\rangle \left\langle e_0\right|
\left\langle s_{\bot }^{iB}\right| \right) U_{BC}^{+}\right\} 
\]

\begin{equation}
=\lambda ^i\left| s^{iB}\right\rangle \left\langle s^{iB}\right| +\left(
1-\lambda ^i\right) \left| s_{\bot }^{iB}\right\rangle \left\langle s_{\bot
}^{iB}\right| .  \label{r28}
\end{equation}

Let us set $\sigma _3^{\left( iB\right) }=\left| s^{iB}\right\rangle
\left\langle s^{iB}\right| -\left| s_{\bot }^{iB}\right\rangle \left\langle
s_{\bot }^{iB}\right| $. Then $\rho _B^i$ has the following form:

\begin{equation}
\rho _B^i=\frac 12\left( I+2\left( \lambda ^i-\frac 12\right) \sigma
_3^{\left( iB\right) }\right) .  \label{r29}
\end{equation}
We can use a completely positive and trace-preserving superoperator$%
\widetilde{V}$ to characterize the operation for particle $B_i$:

\begin{equation}
\widetilde{V}\left( \left| s^{iB}\right\rangle \left\langle s^{iB}\right|
\right) =\frac 12\left( I+a_i\overrightarrow{\sigma }_{i1}\right) ,
\label{r30}
\end{equation}

\begin{equation}
\widetilde{V}\left( \left| s_{\bot }^{iB}\right\rangle \left\langle s_{\bot
}^{iB}\right| \right) =\frac 12\left( I+b_i\overrightarrow{\sigma }%
_{i2}\right) ,  \label{r31}
\end{equation}
where $\overrightarrow{\sigma }_{i1}$ and $\overrightarrow{\sigma }_{i2}$
satisfy $\overrightarrow{\sigma }_{i1}\cdot \overrightarrow{\sigma }_{i1}=I$
and $\overrightarrow{\sigma }_{i2}\cdot \overrightarrow{\sigma }_{i2}=I,$
respectively. The $a_i$ and $b_i$ are real numbers satisfying $\left|
a_i\right| \leq 1$ and $\left| b_i\right| \leq 1$. The superoperator $%
\widetilde{V}$ maps the reduced density matrix $\rho _B^i$ to itself:

\[
\widetilde{V}\left( \rho _B^i\right) =\lambda ^i\widetilde{V}\left( \left|
s^{iB}\right\rangle \left\langle s^{iB}\right| \right) +\left( 1-\lambda
_i\right) \widetilde{V}\left( \left| s_{\bot }^{iB}\right\rangle
\left\langle s_{\bot }^{iB}\right| \right) 
\]

\begin{equation}
=\lambda _i\left[ \frac 12\left( I+a_i\overrightarrow{\sigma }_{i1}\right)
\right] +\left( 1-\lambda _i\right) \left[ \frac 12\left( I+b_i%
\overrightarrow{\sigma }_{i2}\right) \right] .  \label{r32}
\end{equation}
So, we obtain the following relation

\begin{equation}
\lambda _ia_i\overrightarrow{\sigma }_{i1}+\left( 1-\lambda _i\right) b_i%
\overrightarrow{\sigma }_{i2}=2\left( \lambda ^i-\frac 12\right) \sigma
_3^{\left( iB\right) }.  \label{r33}
\end{equation}
Since $\lambda _i$ is unknown, by comparing two sides of the above equation,
we get further relations 
\begin{equation}
\overrightarrow{\sigma }_{i1}=\overrightarrow{\sigma }_{i2}=\sigma
_3^{\left( iB\right) },  \label{r34}
\end{equation}

\begin{equation}
a_i=1,\text{ }b_i=-1.  \label{r35}
\end{equation}
Thus the joint unitary evolution must be

\begin{equation}
U_{BC}\left( \left| s^{iB}\right\rangle \left| e_0\right\rangle \right)
\longrightarrow e^{i\delta _1}\left| s^{iB}\right\rangle \left|
e_1^i\right\rangle ,  \label{r36}
\end{equation}

\begin{equation}
U_{BC}\left( \left| s_{\bot }^{iB}\right\rangle \left| e_0\right\rangle
\right) \longrightarrow e^{i\delta _2}\left| s_{\bot }^{iB}\right\rangle
\left| e_2^i\right\rangle ,  \label{r37}
\end{equation}
where $\left| e_1^i\right\rangle $ and $\left| e_2^i\right\rangle $ are two
normalized states satisfying $\left\langle e_1^i\mid e_1^i\right\rangle =1$
and $\left\langle e_2^i\mid e_2^i\right\rangle =1,$ respectively. After the
joint unitary evolution, the inseparability correlation coefficient of the
system AB is $I_c\left( \rho _{AB}^{\left( i\right) ^{\prime }}\right) =4%
\sqrt{\lambda _i\left( 1-\lambda _i\right) }\left| \left\langle e_1^i\mid
e_2^i\right\rangle \right| .$ To meet the requirement of $I_c=0$, $\left|
e_1^i\right\rangle $ and $\left| e_2^i\right\rangle $ must be orthonormal.
As the final result, we find that $\left| e_1^i\right\rangle $ and $\left|
e_2^i\right\rangle $ are two orthonormal states. Thereby, the reduced
density matrices of the additional system and the particle $B_i$ have the
same form. In essence, a universal broadcasting process of the mixed states
is fulfilled. This result is conflict with the result that noncommuting
mixed states cannot be broadcast\cite{Barnum}.

If one performs a projection measurement on the additional system, an
uncontrolled change of the reduced density matrix of the particle A will
lead to the failure of ideal disentanglement. While, any operation in
quantum mechanics can be represented by a generalized unitary evolution,
together with a measurement. Hence, we complete the proof.

\section{Properties of Inseparability Correlation Coefficient}

In this section, we will embark on analyzing the properties of
inseparability correlation coefficient under some local operations. For $I_c$
coefficient we have:

Prop 1 : $I_c\left( \rho \right) =0$ if $\rho $ is separable. When the given
system is in pure state or Bell mixed state $I_c=0$ can be used as necessary
and sufficient criterion of separability.

Prop 2 : Local unitary operations leave $I_c\left( \rho \right) $ invariant,
i.e. $I_c\left( \rho \right) =I_c\left( U_A\otimes U_B\rho U_A^{+}\otimes
U_B^{+}\right) $.

For Prop 1, we can refer to \cite{Horodecki1} and \cite{account}. Prop 2 is
evident because local unitary transformations can not give rise to the
change of three eigenvalues of the matrix M. However, we are more concerned
with the properties of inseparability correlation coefficient under local
general operations. Does it have the properties similar to the entanglement
measures?

For a qubit system, its density matrix has the following form:

\begin{equation}
\rho =\frac 12\left( I+\overrightarrow{r}\cdot \overrightarrow{\sigma }%
\right) .  \label{r39}
\end{equation}
A general operation maps $\rho $ to $\rho ^{\prime }$ \cite{Chuang}:

\begin{equation}
\rho ^{\prime }=\widetilde{V}\left( \rho \right) =\frac 12\left( I+%
\overrightarrow{q}\cdot \overrightarrow{\sigma }\right) ,  \label{r40}
\end{equation}

\begin{equation}
\overrightarrow{q}=\overrightarrow{\delta }+\overrightarrow{r}\bullet B,
\label{r41}
\end{equation}
where $\overrightarrow{\delta }$ is a constant vector in $R^3,$ and B is a 3$%
\times $3 real matrix. This is an affine map, mapping the Bloch sphere into
itself. Under the action of the superoperator $\widetilde{V}$, the identity
and Pauli operators have the following map relations:

\begin{equation}
\left\{ 
\begin{array}{c}
\widetilde{V}\left( I\right) =I+\overrightarrow{\delta }\cdot 
\overrightarrow{\sigma } \\ 
\widetilde{V}\left( \sigma _i\right) =\sum_{j=1}^3B_{ij}\sigma _j
\end{array}
\right. .  \label{r42}
\end{equation}
For a 2$\times $2 system described by Eq(\ref{r7}), we perform a general
operation on the first subsystem. Here, we only consider the case of $%
\overrightarrow{\delta }=0$, in which adopting the following trick will
excuse us from complicated calculations.

We can introduce an auxiliary Bell mixed state $\rho _B$ corresponding to
the state $\rho _{12}$ described by Eq(\ref{r7}): 
\begin{equation}
\rho _B=\frac 14\left( I\otimes I+\sum_{n,m=1}^3t_{mn}\sigma _m\otimes
\sigma _n\right) .  \label{a11}
\end{equation}
With the limit of $\overrightarrow{\delta }=0$, we have: 
\begin{equation}
T\left( \rho _{12}^{\prime }\right) =B^TT\left( \rho _{12}\right)
=B^TT\left( \rho _B\right) =T\left( \rho _B^{\prime }\right) .  \label{a12}
\end{equation}
Because inseparability correlation coefficient only depends on the matrices
T we can study the varies of Bell mixed state $\rho _B$ instead of the state 
$\rho _{12}$.

With regard to Bell mixed states, their entanglement of formation has been
given by C. H. Bennett et. al \cite{Bennett2}: 
\begin{equation}
E\left( \rho \right) =h\left( f\left( \rho \right) \right) ,  \label{a13}
\end{equation}
where f($\rho $) refers to fully entangled fraction. The function h(f) is
defined as: 
\begin{equation}
h\left( f\right) =\left\{ 
\begin{array}{c}
H\left[ \frac 12+\sqrt{f(1-f)}\right] ,\text{ }\left( f\geq \frac 12\right)
\\ 
0,\text{ }(f<\frac 12)
\end{array}
.\right.  \label{a14}
\end{equation}
Here, $H(x)=-x\log _2x-(1-x)\log _2\left( 1-x\right) .$ In accordance with
Eqs (\ref{a2},\ref{a13},\ref{a14}), we find that E($\rho $) is a
monotonically increasing function of N($\rho $) for the Bell mixed state $%
\rho $ in inseparability correlation region. While, an accepted fact is that
entanglement cannot increase under local operations and classical
communications. Therefore, N($\rho $) cannot increase. As far as the Bell
mixed states in separability correlation region are concerned, N($\rho $)
cannot exceed one under local operations. Thus, we conclude that
inseparability correlation coefficient $I_c\left( \rho \right) $ can not
increase under a class of special local operations ( $\overrightarrow{\delta 
}=0$ ) and classical communications. i. e. :

\begin{equation}
I_c\left( \rho _{12}\right) \geq I_c\left( \widetilde{V}\left( \rho
_{12}\right) \right) =I_c\left( \rho _{12}^{\prime }\right) .  \label{a15}
\end{equation}

In the case of $\overrightarrow{\delta }\neq 0,$ the properties of
inseparability correlation coefficient are not clear yet. We guess, under
more general operations, $I_c\left( \rho \right) $ cannot increase. We will
go on with this problem in our future work.

\section{Discussion and Conclusion}

Under local operations, to achieve general isotropic disentanglement, a
threshold limit has been shown in section 2. A further question is that if
nonlocal operations are allowed, to make reduced density matrices shrink
isotropically, what threshold limit will be attained? S. Ghosh et. al hold
that local disentangling machines are better than nonlocal ones\cite{Ghosh}.
But they do not give a theoretical proof. We think it is still an open
question.

Local cloning operations can be used for broadcasting entanglement\cite
{Buzek} and for disentangling. Recently, the N$\rightarrow $M optimal
universal quantum cloning has been studied \cite{Bruss}\cite{Werner} by
Bruss et. al and Werner, respectively. For the 1$\rightarrow $M-qubit
cloning operations, the optimal isotropic reduction factor is $\eta _{\max }=%
\frac{M+2}{3M}.$ If symmetrically performing local cloning operations on an
unknown two-qubit system, based on the threshold inequality $\eta ^2\leq 
\frac 13,$ one can derive M$\geq $3, i. e. the $1\rightarrow 3$ cloning
operations on each subsystem will disentangle the initial two-qubit system.
Furthermore, let us consider nonlocal cloning cases in which the two-qubit
is cloned as a whole. The maximal reduction factor is $\eta _{\max
}^{^{\prime }}=\frac{M+4}{5M}.$ ( Note that here $\eta ^{^{\prime }}$ is a
reduction factor in 2$\times $2 Hilbert space. The original matrix T is
shrunk by $\eta ^{^{\prime }}.$ ) Using the method similar to Section 2, we
can conclude that : under this type of operations, the necessary and
sufficient condition realizing universal disentanglement is $\eta ^{\prime
}\leq \frac 13.$Therefore, the result of $M\geq 6$ is derived, i. e. under
nonlocal cloning operations, the 1$\rightarrow $6-pair entanglement
broadcast will disentangle any two-qubit system. By contrast, we can
conclude that quantum inseparability can be copied better by a nonlocal
copier than by a local copier.

In the quantum information field, the quantitative degree of entanglement is
attracting more and more attention. In recent years, several entanglement
measures have been proposed\cite{Bennett2}\cite{Vedral}. Bennett et. al use
Bell singlet state as a quantitative standard and introduce the concepts of
entanglement of formation and entanglement of distillation \cite{Bennett2},
which characterize two major aspects about given ensemble of entanglement.
Vedral et. al define entanglement measure as the minimum distance between
given density matrix and the subset containing all disentangled states \cite
{Vedral}. This scientific description provides a clear distinction between
quantum entanglement and classical correlations. Inseparability correlation
coefficient, which may act as a sufficient criterion for entanglement, can
not be used to quantify of entanglement. Because the entanglement hidden in
the mixed states is not only determined by the matrix T($\rho $) but also
embodied in the relations between $\overrightarrow{r}$, $\overrightarrow{s}$
and T($\rho $)\cite{Horodecki3}. Nevertheless, we believe that there exist
close connections among the matrix T($\rho $), Bloch vectors of the
subsystems and all kinds of definitions on entanglement measure. Studying
the relationship is still a meaningful work.

In conclusion, we study local disentanglement processes in two-qubit system,
and give the conditions realizing ideal disentanglement provided that some
information on quantum states is known. Furthermore, under some local
operations, we analyze the behaviors of inseparability correlation
coefficient. We believe that the results of the present paper can help in
deeper understanding of the connection between entanglement and
disentanglement.

Note added: Recently, we found that Ghosh et al. replaced their paper\cite
{Ghosh}, in which the threshold inequality $\eta _1\eta _2\leq \frac 13$ is
also obtained.

\vskip 5mm

This project was supported by the National Nature Science Foundation of
China.

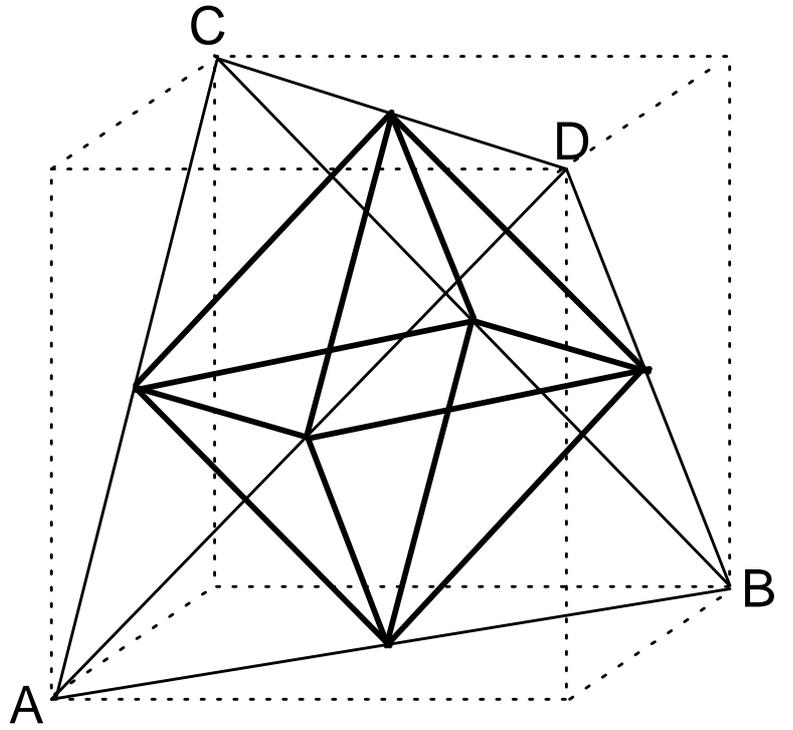
\begin{figure}[tbp]
\caption{Geometrical representation of characteristic vectors of the
matrices T, the tetrahedron $\widetilde{T}$ represents all states in
two-qubit system and the bold-line-coutoured octahedron contains the subset
of separable states. Here, A=(-1, -1, -1), B=(1, 1, -1), C=(1, -1, 1),
D=(-1, 1, 1).}
\label{fg:Fig.1}
\end{figure}

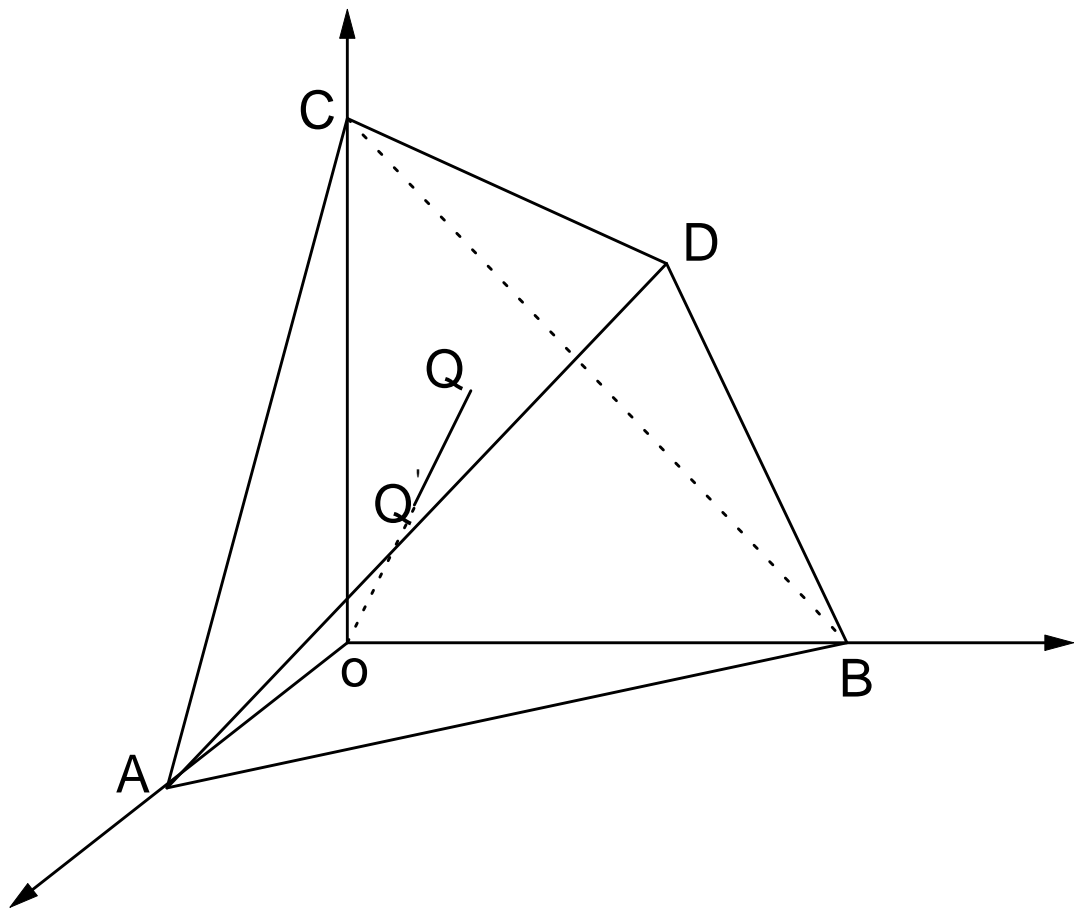
\begin{figure}[tbp]
\caption{The space of characteristic vectors of the matrices M, the
tetrahedron $\widetilde{T}_{oABC}$ contains characteristic vectors of all
separable states. Here, Q is a dot in inseparability correlation region $%
\widetilde{T}_{ABCD}$. Line segment oQ intersects plane ABC at dot Q'.}
\label{fg:Fig.2}
\end{figure}


\begin{references}
\bibitem{account}  For any a 2$\times $2 system, it is easy to verify that
separable and inseparable pure states satisfy N($\rho $)=1 and N($\rho $)%
\mbox{$>$} 1 respectively. Therefore, separability correlation region $%
\widetilde{T}_{oABC}$ cannot contain any characteristic vector $%
\overrightarrow{m}$ of pure entangled states.
\end{references}
\end{document}